\pdfoutput=1

\documentclass[fleqn,usenatbib]{mnras}

\usepackage{newtxtext,newtxmath}

\usepackage[T1]{fontenc}

\DeclareRobustCommand{\VAN}[3]{#2}
\let\VANthebibliography\thebibliography
\def\thebibliography{\DeclareRobustCommand{\VAN}[3]{##3}\VANthebibliography}

\usepackage{graphicx}	
\usepackage{amsmath}	

\usepackage{amssymb}	

\title{Doppler Signature of a Possible Termination Shock in an Off-Limb Solar Flare}

\author[R. J. French et al.]{
Ryan J. French,$^{1}$
Sijie Yu,$^{2}$
Bin Chen,$^{2}$
Chengcai Shen$^{3}$
and Sarah A. Matthews$^{4}$
\\
$^{1}$National Solar Observatory, 3665 Discovery Drive, Boulder, CO 80303, USA \\
$^{2}$Center for Solar-Terrestrial Research, New Jersey Institute of Technology, 3L King Jr. Blvd., Newark, NJ 07102-1982, USA\\
$^{3}$Harvard-Smithsonian Center for Astrophysics, 60 Garden St., Cambridge, MA 02138, USA \\
$^{4}$University College London, Mullard Space Science Laboratory, Holmbury St. Mary, Dorking, Surrey, RH5 6NT, UK
}

\date{Accepted to MNRAS, Feb 6th 2024}

\pubyear{2024}

\begin{document}
\label{firstpage}
\pagerange{\pageref{firstpage}--\pageref{lastpage}}
\maketitle

\begin{abstract}
We report striking Doppler velocity gradients observed during the well-observed September 10th 2017 solar flare, and argue that they are consistent with the presence of an above-the-looptop termination shock beneath the flare current sheet. Observations from the Hinode Extreme-ultraviolet Imaging Spectrometer (EIS) measure plasma sheet Doppler shifts up to 35 km/s during the late-phase of the event. By comparing these line-of-sight flows with plane-of-sky measurements, we calculate total velocity downflows of 200+ km/s, orientated $\approx 6-10^\circ$ out of the plane of sky. The observed velocities drop rapidly at the base of the hot plasma sheet seen in extreme ultraviolet, consistent with simulated velocity profiles predicted by our 2.5-D magnetohydrodynamics model that features a termination shock at the same location. Finally,
the striking velocity deceleration aligns spatially with the suppression of Fe XXIV non-thermal velocities, and a 35--50 keV hard X-ray looptop source observed by the Reuven Ramaty High Energy Solar Spectroscopic Imager (RHESSI). Together, these observations are consistent with the presence of a possible termination shock within the X8.2-class solar flare.

\end{abstract}




\section{Introduction}

The standard model of eruptive solar flares, also known as the CSHKP model after the original works from \citet{Carmichael1964,Sturrock1968,Hirayama1974,Kopp1976}, is able to explain many observable phenomena in solar flares. In this model, magnetic reconnection is driven by the inflow of oppositely oriented magnetic fields beneath an erupting magnetic flux rope. Within this reconnection site, magnetic free energy is converted into plasma heating, particle acceleration, and electromagnetic radiation. Newly formed magnetic fields are ejected from either end of the current sheet site, contributing towards the erupting flux rope (and associated coronal mass ejection) above, and forming hot flare loops, below. 

However, the standard model is unable to account for some of the higher-energy features observed in flares. In particular, it does not address the specific physical mechanisms through which magnetic energy is converted into the kinetic energy of accelerated particles observed in solar flares.
In recent years, proposed additions to the simplified 2-dimensional standard flare model have been developed to explain observations of high energy particles. The cartoon in Figure 3C of \citet{Chen2015} presents such a scenario after \citet{Masuda1994}, incorporating hard X-ray (HXR) and radio observations into the standard model configuration, explained by the presence of a termination shock. In solar flares, a termination shock will persist as super-magnetosonic reconnection outflows from the reconnection site, collide with the dense post-reconnected flare loops. Termination shocks of similar magnetic geometry are also believed to occur in other astrophysical plasmas, such as in supernova remnants \citep[e.g.][]{Miles2009}.

However, convincing evidence of solar flare termination shocks are comparatively rare.  
\citet{Chen2015} and \citet{Chen2019} use high-cadence radio imaging spectroscopy to reveal the morphology and dynamics of a termination shock accelerating solar flare energetic electrons, using observations of a C1.9 class eruptive flare from the Karl G. Jansky Very Large Array (VLA). Work from \citet{Luo2021} finds similar results, with VLA observations of a much larger M8.4-class flare.
The synergy of ultraviolet and X-ray observations have also been used to deduct the likely presence of a termination shock, in work from \citet{Polito2018}. \citet{Polito2018} detect large (200+ km/s) Doppler shifts in the high temperature Fe XXI line with the Interface Region Imaging Spectrograph \citep[IRIS, ][]{DePontieu2014}. These Doppler shifts, observed in an on-disk X1-class flare, were found to diverge above the flare loop arcade, aligning with a high energy 35--70 keV HXR source observed by the Reuven Ramaty High Energy Solar Spectroscopic Imager \citep[RHESSI, ][]{Lin2002}. The authors attributed these observations to deflecting flows consistent with the existence of a termination shock above the looptop. These studies all provide observational evidence that strongly supports the existence of a termination shock, through measurements of ion dynamics and electron populations.
Simulations have also been able to produce termination shocks within solar flares, first in 2D \citep{Forbes1986,Takasao2015,Shen2018} and more recently 3D \citep{Shen2022,Shibata2023}.

According to predicted termination shock geometry, solar flare termination shocks are sat within hot cusp-like features, above the bright flare loops, but below the current sheet base. These hot cusp shapes are observed frequently in solar flares at the limb of the Sun, easy to spot in the hot 131 and 94 \AA\ channels of the Solar Dynamics Observatory's Atmospheric Imaging Assembly \citep[SDO/AIA, ][]{Lemen2012}. The tip of the cusp meets where we expect the base of the current sheet to be, but observations of visible heated plasma around the current sheet (known as a plasma sheet), are far more infrequent. Because of their infrequency, spectroscopic measurements of the plasma sheet and cusp region in (E)UV are even rarer. Only a handful of spectroscopic observations of plasma sheets or cusp regions exist for solar flares, including the famous September 10th 2017 flare. Many facets of this flare have been studied, including dynamics of the hot, long-lived plasma sheet \citep[e.g, ][]{Warren2018, Cheng2018, Longcope2018, French2019, Chen2020a}, flare loop-top region \citep{Chen2020a, Fleishman2020, Yu2020}, and flare arcade to the south \citep{Reeves2020} that is associated with the 3D geometry of the erupting flux rope \citep{Chen2020b}.
In this study, we revisit spectral observations from the Hinode EUV Imaging Spectrometer \citep[EIS, ][]{Culhane2007} of the September 10th 2017 flare, in particular the late phase observations presented by \citet{French2020}. We combine EIS Doppler measurements along the plasma sheet with plane-of-sky (POS) EUV downflows detected within the same region and time period by \citet{Yu2020}. Together, the Doppler (line-of-sight, LOS) and POS downflows provide a 3D measurement of the velocity profile, which we compare with velocities predicted by termination shock simulations.

\section{Summary of Observations}

\begin{figure*}
\centering
\includegraphics[width=17.8cm]{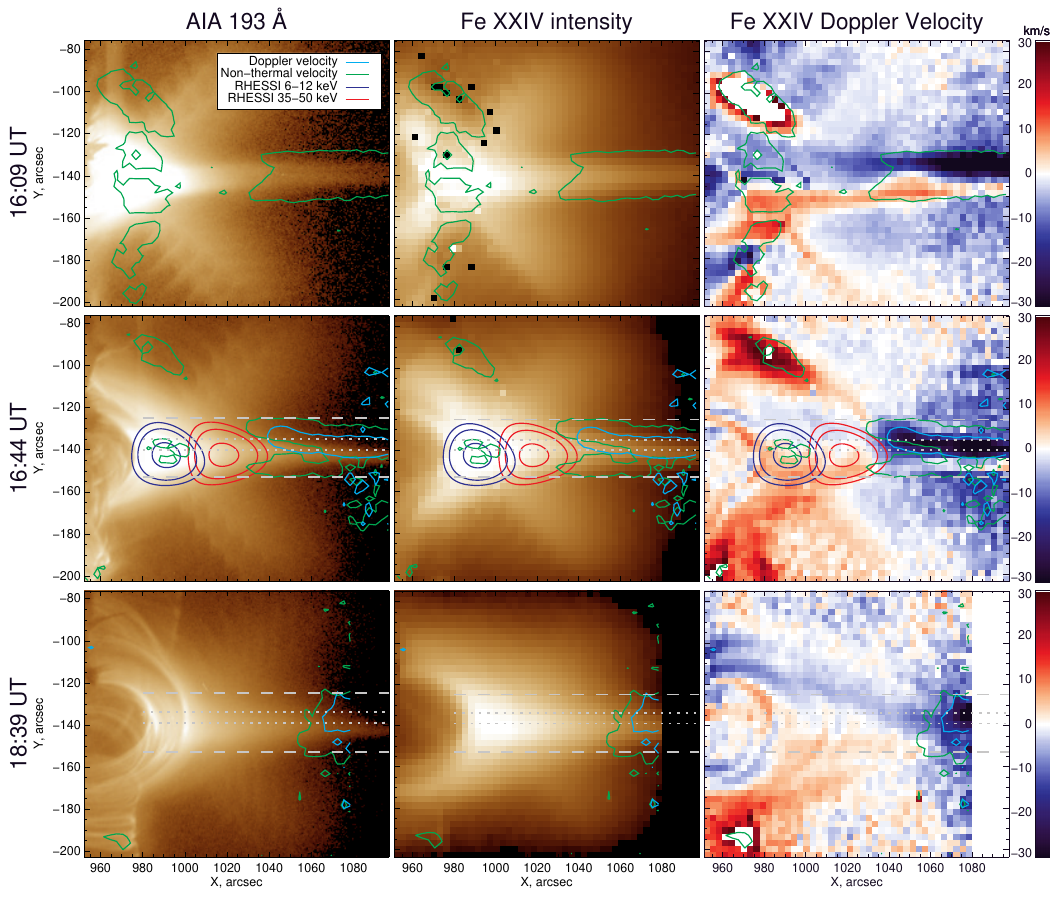}
\caption{Left column: AIA 193 \AA\ images at 16:09, 16:44 and 18:39 UT. Times are consistent across each row. Middle column: Hinode-EIS Fe XXIV 192.04 \AA\ intensity. Right column: Hinode-EIS Fe XXIV 192.04 \AA\ Doppler velocity. Green and cyan contours provide spatial locations of highest non-thermal velocity and Doppler velocity respectively. Doppler velocity contours are set at $-$16 km/s, and non-thermal velocity contours 100, 70, \& 60 km/s for 16:09, 16:44 and 18:39 UT respectively. Doppler contours are not included in the 16:09 UT frame, due to PSF effects. For the 16:44 UT maps, RHESSI contours of 6--12 keV and 35--50 keV HXR emission are also included as blue and red contours respectively. The four gray horizontal-dashed lines at 16:44 and 18:39 UT mark the \textit{Narrow} (short dashes) and \textit{Wide} (long dashes) plasma sheet cross-sections analyzed in Figure \ref{fig:profile}.
}
\label{fig:images}
\end{figure*}

The X8.2-class September 10th 2017 flare erupted over the western limb from AR 12673. Due to its position on the Sun and within the solar cycle, the September 10th 2017 flare was observed across the electromagnetic spectrum by a fleet of space and ground-based instruments, making it the most published solar flare of all time \citep{French2022}.

The first column of Figure \ref{fig:images} shows AIA 193 \AA\ emission of the flare over the west limb, at snapshots coinciding with the start of three Hinode EIS rasters. The AIA 193 \AA\ broadband filter contains two strong spectral lines, the cool Fe XII 193 line (peaking at 1.6 MK) and hot Fe XXIV 192.04 \AA\ line (peaking at 18 MK). The hot Fe XXIV line is only produced in the strongest of flares, so not common in AIA 193 \AA\ images. The three rows of Figure \ref{fig:images} show the evolution of AIA 193 \AA\ emission, with growth of the flare loops and hot cusp region, and persistence of the hot, horizontal plasma sheet.

Hinode-EIS observed the September 10th 2017 flare for long periods of its duration, including from flare onset at 15:44 UT, through the peak at 16:07 UT, up to 16:52 UT. Observations continued later in the flare's evolution, from 18:39 UT to 19:31 UT. EIS observed a 2\arcsec\ slit-width raster scan, rastering from west to east with a cadence of 8 minutes 52 seconds. EIS measured spectra of multiple spectral lines, including the hot Fe XXIV 192.04 \AA\ line included in the broadband AIA 193 \AA\ window. The hot formation temperature of Fe XXIV renders it useful for studies of hot flaring plasma. The second column of Figure \ref{fig:images} presents intensity measurements of EIS Fe XXIV 192.04 \AA\ aligned with AIA, obtained by fitting a single Gaussian curve to 1x3 binned data (to create square 3x3\arcsec pixels). The three EIS intensity maps presented in Figure \ref{fig:images} were chosen to show the first EIS raster of the plasma sheet at 16:09 UT, the final EIS raster of the first observing sequence (16:44 UT), and first EIS raster of the continued observations at 18:39 UT. That is, EIS was collecting 9 minute rasters between the first two frames of Figure \ref{fig:images}, but with a data gap between the second and third frame. In these images, the solar limb is just off the left edge of the image field-of-view (FOV). There are clear similarities between the EIS 192.04 \AA\ and AIA 193 \AA\ emission, specifically in the presence of the hot plasma sheet and above-looptop cusp region. All EIS maps have been cropped from their initial FOV. Due to a shift in the EIS pointing from 18:39 UT onwards, the right edge of the 18:39 UT FOV sits within the plotted region.

Gaussian fitting of the Fe XXIV emission line also provides diagnostics of Doppler velocity and non-thermal velocity (excess line broadening beyond the thermal width) for the flaring-temperature plasma. 
The third column of Figure \ref{fig:images} show maps of EIS Fe XXIV Doppler velocity, aligned to the other intensity panels in the figure. Due to the shape of the point-spread function (PSF), care is needed when interpreting off-limb Doppler measurements, which can generate artifacts along the slit direction. 
These artifacts are most prominent at sharp boundaries between high and low intensity regions, such as that between the bright off-limb features and the dark background of space. This PSF has a stark effect in the early EIS Doppler velocity maps of the flare examined here, within the bright plasma sheet.
In Figure \ref{fig:images}, the slit orientation is top to bottom, rastering from right to left. In the bright plasma sheet region, photons from bright pixels can contaminate the wings of adjacent pixels, creating an artificial Doppler signature. This PSF effect is discussed in detail by \citet{Warren2018}, and easily seen in the 16:09 UT Doppler velocity frame in Figure \ref{fig:images}. In this frame, horizontal streaks of strong parallel red and blue shifts are seen along the plasma sheets. These Doppler measurements are not real, and an artifact of the EIS PSF. Later in the flares evolution, growth of the cusp region around the plasma sheet removes the sharp intensity transition, and therefore reduces influence of the PSF on the Doppler maps. For these later Doppler maps \citep[as presented in][]{French2020}, we can reliably extract LOS velocity information from the Doppler profiles. Examining the Doppler maps at 16:44 and 18:39 in Figure \ref{fig:images}, this study takes particular interest at the strong blue shift signal within the plasma sheet region towards the right of the frame, and strong velocity gradient at the base of the plasma sheet. For these two later EIS rasters, blue-shift contours (at $-$16 km/s) are overlaid onto the intensity maps in left and center panels (in cyan). We use the standard definition for blue shift in this study, defined as velocities flowing \textit{towards} the observer. Blue shift velocities at the base of the plasma sheet, where plasma is flowing towards the Sun, immediately indicate that the plasma sheet is tilted slightly away from us, out of the POS above the limb. This is consistent with the known geometry of the September 10th 2017 flare \citep[e.g.][]{Chen2020b}.

Also included within every panel in Figure \ref{fig:images} are green contours of EIS 192.04 \AA\ non-thermal velocity. Values of 100, 70 and 60 km/s are chosen for the three presented EIS times respectively. EIS non-thermal velocity measurements for this flare have been examined previously, e.g. \citet{Warren2018,Longcope2018} for the earlier EIS rasters, and \citet{French2020} for the later EIS rasters. The non-thermal velocities have been interpreted in these studies as macroscopic turbulence along the LOS, as expected from reconnection-induced turbulence within the plasma sheet.

Intermittent measurements of HXRs were also observed by the RHESSI for this flare, which had four operational detectors at the time (1, 3, 6, 8). For the 16:44 UT panels in Figure \ref{fig:images}, RHESSI 6-12 and 35-50 keV emission are overplotted with blue and red contours respectively. The RHESSI HXR images are the same values presented in Figure 2e of \citep{Gary2018}, and were reconstructed using the clean algorithm with data from detectors 3, 6 and 8. The 6--12 keV signal, consisting primarily of thermal emission, is located at the top of the flare loops. The 35--50 keV signal, however, containing mainly non-thermal emission, peaks higher above the limb at the intersection between the hot flare cusp and plasma sheet. Unfortunately, RHESSI was not observing during the later 18:39 UT period.

\section{velocity gradients along the plasma sheet}
\subsection{Line-of-Sight Velocities}

\begin{figure*}
\centering
\includegraphics[width=15.5cm]{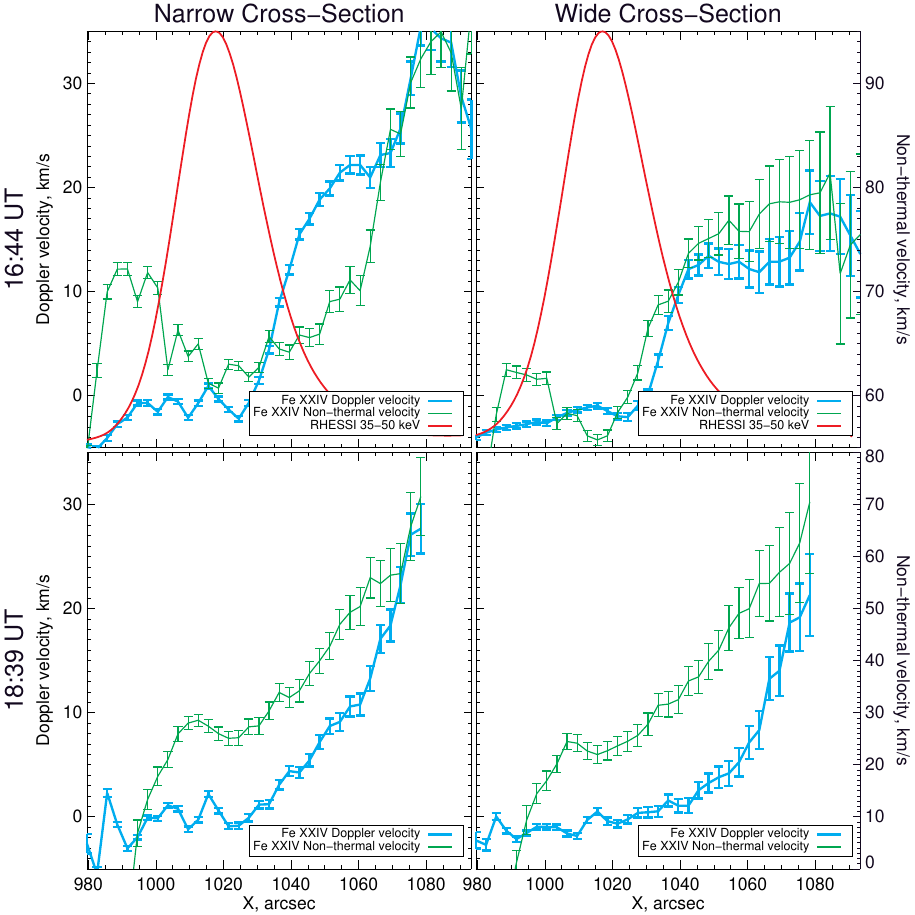}
\caption{Plane-of-sky gradients of EIS 192.04 \AA\ Doppler and non-thermal velocity, relative to 35-50 keV RHESSI emission. The RHESSI emission curve is normalised to the maximum/minimum y axis values. The two columns present the cross-section of these values along the narrow and wide cross-sections presented in Figure \ref{fig:images}, respectively. The top and bottom row each show the same cross-sections at 16:44 and 18:39 UT.
}
\label{fig:profile}
\end{figure*}  

Dashed horizontal gray lines in Figure \ref{fig:images} mark the locations of two cross-sections along the plasma sheet. The outer dashed lines mark the \textit{Wide} cross-section, which encompasses all EUV emission from the current sheet. The inner dashed lines mark the location of the \textit{Narrow} cross-section, centered within the strong blue-shift signal in the plasma sheet. Along these regions, we take the cross-section of EIS Fe XXIV Doppler and non-thermal velocities, with RHESSI 35-50 keV emission. Values are averaged in the y-direction, across the cross-section region width.

Figure \ref{fig:profile} plots both the \textit{Narrow} and \textit{Wide} cross-sections for the 16:44 and 18:39 UT time frames. Doppler velocity values are plotted on the left y-axis, and non-thermal velocity values on the right y-axis. The Doppler velocity shows blue shift, but we plot the magnitude of the blue shift for ease of comparison to the non-thermal velocity trends.

All four panels demonstrate a strong gradient in both Doppler and non-thermal velocities, suggesting a rapid deceleration of plasma as it approaches around $x=1020\arcsec$. Both EIS Fe XXIV parameters then experience another gradual or rapid decline below $x=1020\arcsec$ at lower altitudes, but it is this first drop and local minimum we are most interested in here.
The gradient is most pronounced in the narrow cross-section, but still visible along the wider region cross-section. Both Doppler and non-thermal velocity demonstrate similar trends, especially in the wide cross-section, where the rate of velocity decrease match for both time periods, (the lines are close to parallel with the margin of error). The gradients between each parameter differ more-so along the narrow-cross-section, resulting from the fact that maximum Doppler and non-thermal velocities are positioned slightly differently at a given width across the plasma sheet (Doppler velocities are centered off-center across the plasma-sheet, whilst the non-thermal velocity shows a wider, bifurcated pattern). Within the narrow cross-section, blue shift Doppler velocities decrease from 35 to 0 km/s in under 60\arcsec\ down the plasma sheet, during both raster times.

At 16:44 UT, RHESSI observations are available for comparison. At this time, the strong EIS Fe XXIV velocity gradients disappear at the same location as peak high energy 35-50 keV RHESSI emission. A sudden velocity drop and a HXR source located above the flare arcade are independently two key signatures expected from the presence of a termination shock, which we see here at both $\approx$ 1 $\&$ 3 hours into the flare evolution respectively.  

\subsection{Plane-of-Sky velocities}

\begin{figure*}
\centering
\includegraphics[width=16cm]{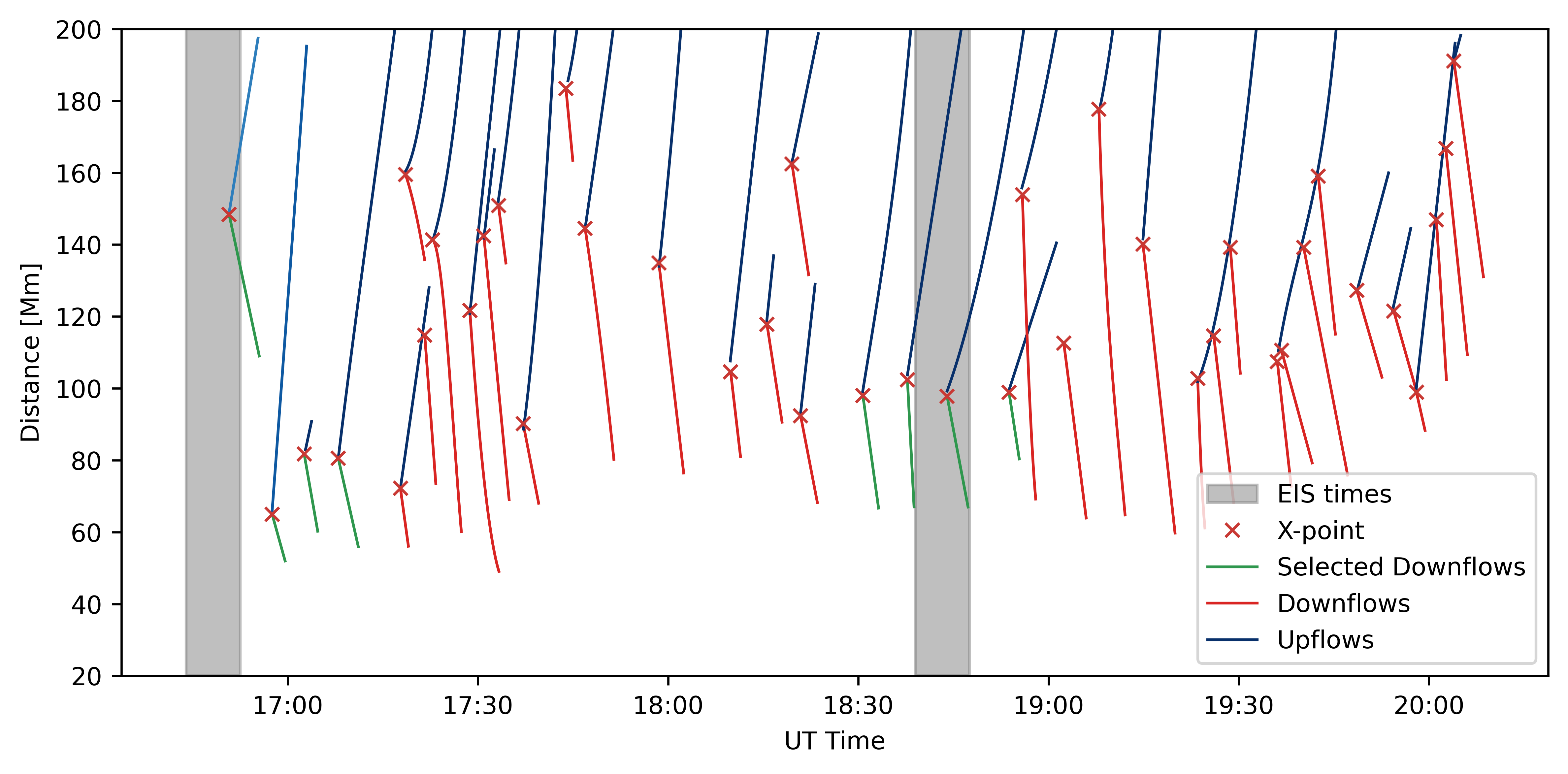}
\caption{Plasma sheet downflows and upflows along the September 10th 2017 flare, detected in difference imaging of AIA 131 \AA\ observations, \citep[modified from Figure 8C in][]{Yu2020}. Red lines mark upflows, red downflows, and green selected downflows near EIS raster times. X's mark the X-points identified between upflow/downflow pairs. Plane-of-sky velocities are distributed between 100-900 km/s, with an average of 250 km/s.
}
\label{fig:downflows}
\end{figure*}

In \citet{Yu2020}, the authors use running difference analysis of 131 \AA\ images to detect faint downflows and upflows along the September 10th 2017 plasma sheet. In doing this, a number of reconnection X-points are located along the plasma sheet structure during this later-phase of the flare.
Figure \ref{fig:downflows} shows a modified version of Figure 8c from \citet{Yu2020}, displaying the timings and height above the limb of all detected downflows, upflows and X-points.
Upflows and downflows are plotted in blue and red respectively, with the exception of the four downflows closest to each EIS raster time, which are plotted in green. EIS raster times are marked with gray dashed lines.

By tracking these plasma flows, we can measure the plasma velocity within the POS, complementary to the LOS velocities measured by EIS Fe XXIV Doppler values. The AIA 131 \AA\ channel samples the hot Fe XXI line, which has a peak temperature sensitivity of 10 MK. This is cooler than the Fe XXIV line measured by EIS, which has a peak temperature sensitivity of 18 MK. Despite this temperature difference, both instruments are observing hot flaring plasma within the plasma sheet, and thus an empirical comparison of POS and LOS velocities are valid.

The downflows plotted in Figure \ref{fig:downflows} have POS velocities ranging from 100-900 km/s, with an average of 250 km/s. Using basic right-angle trigonometry, we can compare the observed 250 km/s POS velocity and 35 km/s Doppler velocity to provide an initial estimate for the magnitude of the total plasma downflow velocity vector. A trigonometric comparison with these values yields a plasma sheet orientation 8$^\circ$ away from the POS, with 99\% of the full velocity vector captured within the observed POS downflows. To take a deeper investigation into the velocity vector and how it changes along the plasma sheet, we can compare observations to simulated plasma velocities from MHD simulations.

\section{Comparison to Simulations}
\begin{figure*}
\centering
\includegraphics[width=15cm]{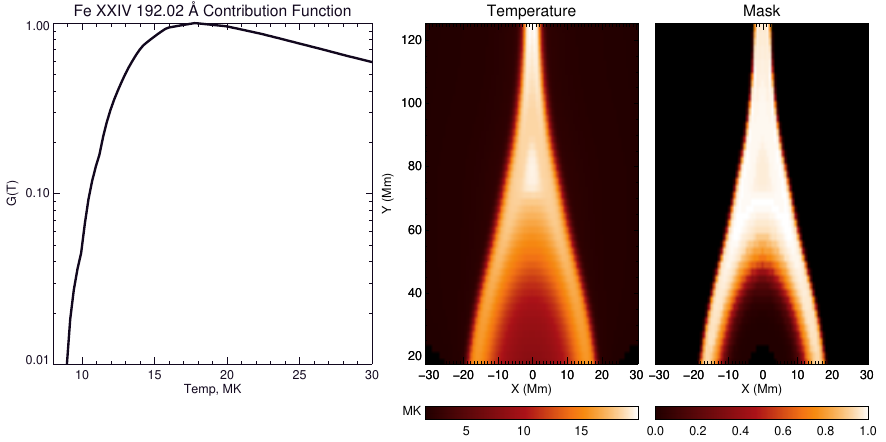}
\caption{Left: Contribution function for Fe XXIV 192.02 \AA. Center: Temperature map of the MHD model frame presented in Figure \ref{fig:velocities}(top). Right: Fe XXIV temperature mask for temperature map in center frame.
}
\label{fig:mask}
\end{figure*} 

\subsection{Model Scaling and Temperature Mask}

We compare observed velocities in the LOS and POS with the simulated velocity profile of a termination shock, utilizing the 2.5D Athena flare model \citep{Shen2018}. The model is dimensionless, and so scaling of the simulation output is needed to compare with observations. Scales for four primary characteristics are defined, and used to calculate the remaining dependent characteristics. The first three primary characteristics are plasma beta ($=0.1$), ambient temperature ($=2$ MK) and magnetic field scaling (40 G), set as standard values for the flaring corona. The fourth primary scaling characteristic is length scale, which we set to match the limb to looptop distance between the simulation and observations, independently for the 16:44 and 18:39 UT time periods of the September 10th 2017 flare. The dependent characteristics, including pressure, density, temperature, current, time and velocity, are scaled relative to the primary ones. 

As a 2.5D flare simulation, parameters are all symmetric along the LOS axis. 
To more accurately compare simulation LOS velocities with those measured in the hot Fe XXIV line by EIS, we apply a temperature scaling mask to the simulation output. Figure \ref{fig:mask} (left) shows the contribution function of the Fe XXIV 192.04 \AA\ line, normalized to the maximum sensitivity of the line. By cross-referencing the temperature sensitivity to the temperature map output by the simulation (as an example shows in the center panel of Figure \ref{fig:mask}), we can create a scaling mask to determine where in the simulation map contains temperatures analogous to those observed by Fe XXIV. This is a quick empirical method to remove flows of cooler plasma from the velocity comparison to observations. The right panel of Figure \ref{fig:mask} shows an example temperature mask for one simulation frame, essentially removing the contribution of pixels outside of the plasma sheet, cusp, or flare loops.

\subsection{Line-of-Sight Comparison}

Given the geometry of the flare \citep[see ][for simualtion information]{Shen2018}, the flare is positioned in the POS, with plasma downflows primary in this plane. To replicate the LOS velocities observed by EIS, we \textit{tilt} the simulation map out of the POS, simultaneously scaling the LOS and POS velocity components by a small angle. Given the small angles involved, it is not required to alter the POS length scales during this process, given that an angle of $10^\circ$ changes the length scale by $<1\%$.

\begin{figure*}
\centering
\includegraphics[width=17.8cm]{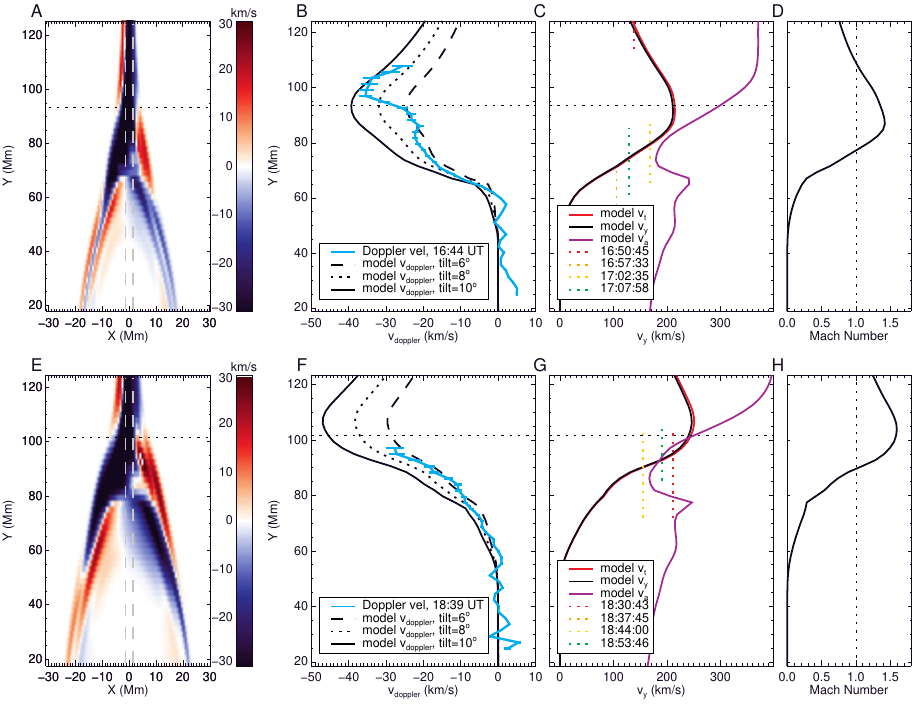}
\caption{Comparison of line-of-sight (LOS) and plane-of-sky (POS) velocities between a simulated solar flare (2.5 ATHENA model) and observations. The top and bottom row present a comparison to the 16:44 and 18:39 UT time periods of the September 10th 2017 flare respectively. The first column presents the LOS velocity map for the flare simulation, tilted 10$^\circ$ out of the POS and with a temperature mask applied (to replicate Fe XXIV emission). The vertical dashed lines mark the location of the plasma sheet cross-section, and horizontal dashed line the termination shock location, determined by the local minimum in the divergence of the velocity field. The second column plots cross-sections along the plasma sheet, comparing simulation (v\textsubscript{doppler}) and observed (EIS Fe XXIV Doppler) LOS velocities. Simulation velocities are plotted at different tilt angles from the LOS. The third column presents the cross-section of velocity down the plasma sheet (in the POS), alongside the total simulation velocity and Alfv\'en speed. The Alfv\'en speed is measured on the outside of the current sheet. The height and speeds of observed POS downflows (from AIA 131 \AA\ observations) are also plotted for comparison. The fourth column plots the cross-section of the Mach number along the model current sheet.
}
\label{fig:velocities}
\end{figure*}

Panel A of Figure \ref{fig:velocities} shows the LOS velocity map of a simulation frame late into the impulsive phase, with an implemented tilt of 10$^\circ$ away from the POS. The Fe XXIV temperature mask has been applied to this map, to empirically replicate the Fe XXIV Doppler velocity observed by EIS. For this simulation frame, we apply a length scaling to create a flare size close the September 10th 2017 flare at the time of the 16:44 UT EIS raster. As described earlier in this section, dependent characteristics (including the characteristic velocity) are then set relative to this length scale and other primary characteristics. This model frame has a termination shock located at the horizontal gray dashed line ($\approx$95 Mm above the limb), determined by a minimum in the divergence of velocity field \citep{Shen2018}.

Qualitatively comparing the simulated LOS velocities in Figure \ref{fig:velocities}A with the EIS Fe XXIV 192.04 \AA\ Doppler maps in Figure \ref{fig:images}, we immediately notice the strong blue shift along the plasma sheet. To compare the two quantitatively, Figure \ref{fig:velocities}B plots cross-sections of LOS velocity for both the simulation and EIS observations. The solid black lines show the cross-section of simulated LOS velocity, taken between the vertical gray dashed lined in panel A. We plot this velocity cross-section for three different tilts, of angles 6 to 10$^\circ$ away from the POS. The cyan curve plots the same EIS Doppler \textit{narrow} cross-section plotted in Figure \ref{fig:profile}. Please note that the axes have been flipped from this earlier Figure \ref{fig:profile}, with distance plotted along the y-axis, and units of arcseconds converted to Mm. 

Figure \ref{fig:velocities}B shows that the observed EIS velocity profile aligns nicely with the simulated Doppler velocities (v\textsubscript{doppler}) of tilts -6 to -10$^\circ$, matching the gradient along the plasma sheet expected from the presence of a termination shock. This comparison was completed for multiple frames from the simulation evolution, with similar results. All of these frames included a termination shock at the base of the plasma sheet.

The second row of Figure \ref{fig:velocities} repeats analysis of the first column, instead comparing the 18:39 UT EIS velocity observations to another simulation frame, later in the simulated flare evolution. To match up the length scales between the two, the characteristic length variable is increased by 25\% from the simulation frame analyzed in the first row, to account for the growth in the flare arcade observed in reality (which is not present over the shorter timescales of the simulation).

Given the shift in the FOV between the 16:44 and 18:39 UT, we do not capture the velocity profile at such high altitudes in the plasma sheet as at 18.39 UT as we did at 16:44 UT. Without observing the height of maximum velocity at this time, a comparison to simulations contains slightly more uncertainty than at earlier times -- with just the rising velocity at lower altitudes captured. Nonetheless, the velocity gradient at the observed section of the plasma sheet aligns nicely with the simulation within the same angle tilt range. Subsequent EIS rasters beyond 18:39 UT contain less and less of the velocity profile, as the loop system continues to rise out of the EIS FOV.

\subsection{Plane-of-Sky Comparison}
In order to fully compare the observed and simulated velocity field, we must also consider the POS velocity. Panel Figure \ref{fig:velocities}C and G compare velocities in the POS. The solid black line shows the simulated velocity along the same plasma sheet cross-section (vertical gray lines in panel A), plotting velocity in the \textit{Y} direction instead of LOS direction (shown in panel B). A tilt of 10$^\circ$ is used to calculate this $v_y$ profile. The solid red line shows the total velocity magnitude along the same slit, containing both the $v_y$ and LOS components of the simulation plasma sheet velocity. Due to the small angle involved ($\cos{[10^\circ]}\approx98.5\%$), the total velocity magnitude is only marginally larger than the $v_y$ component. Similarly to the LOS velocity, the strong velocity gradient, reaching a minimum at the termination shock in the simulation, is also present in the POS velocity.

Also plotted for reference is the Alv\'en velocity ($v_a$) adjacent to the current sheet cross-section (colored magenta). In the adjacent panel, we plot the cross-section of the Mach number along the current sheet, $M = v_t/(c_s^2 + v_a^2)^{0.5}$, where $v_t$ is total velocity and $c_s$ the local sound speed. Comparing current sheet downflow velocities with $v_a$ and $M$, we see plasma velocities start to decrease at the termination shock location, at an altitude of decreasing $v_a$ and $M>1$. $v_a$ then reaches a local minimum as $M$ falls below unity. Although the model provides a lower $v_a$ than we might initially expect, it is important to note that $v_a$ is larger outside of the current sheet than within it. Crucially, despite these lower values at the direct edge of the current sheet, the Mach number remains greater than unity (as required to create the presence of a shock).

To compare simulated POS velocities with observations, we overplot in Figure \ref{fig:velocities}C and G the velocity and heights of plasma sheet downflows presented in Figure \ref{fig:downflows}. We use only the four downflows with times closest to each EIS raster. These downflows are plotted in Figure \ref{fig:velocities} as vertical dashed lines, with a color corresponding to the downflow time. For \ref{fig:velocities}C, where comparison is made to the 16:44 UT time period, all four observed downflows intersect the simulated velocity profile. The same goes for the 18:39 UT comparison in \ref{fig:velocities}G, with the exception of a single, much faster downflow, which is off the scale. The intersection of the observed downflows lines and simulated velocities demonstrate that the simulation, with a velocity profile determined by the presence of a termination shock, is successfully able to replicate both LOS and POS observed velocities. It is important to note that the absence of visible downflows at certain altitudes in AIA 131 \AA\ observations does not necessitate that downflows are not present at these locations and times. As determined by the standard model, downflows are present continuously, with only narrow windows of flowing plasma visible. With measured downflows of 100-170 and 150-210 km/s present around our two time windows (16:44 and 18:39 UT), the simulation comparison would suggest that faster downflows are present, but not observed. The simulations indicate these could be as high as 210 and 240 km/s for each time frame respectively.

\section{Conclusions}

In this work, we have analyzed LOS and POS velocities observed along the plasma sheet of an off-limb flare, at two time periods (16:44 and 18:39 UT) during the flare's late phase. By comparing LOS velocities of 35 km/s (from EIS Fe XXIV 192.04 \AA Doppler observations) with POS velocities of 100-210 km/s (from AIA 131 \AA\ difference imaging), we deduce total velocities marginally higher than the POS measurements, and a flare configuration oriented 6-10$^\circ$ out of the POS. Although plasma downflows are continuous, our detections of them are not. It is therefore probable that further downflows, both faster and slower than those observed, are also present, but do not create the temperature or density contrast necessary to detect in EUV images.

2.5D ATHENA simulations of an eruptive flare are able to replicate velocity magnitudes and gradients in the LOS and POS, through the presence of a termination shock at the base of the plasma sheet. Along the plasma sheet, the simulation also suggests that downflow velocities are higher than we are able to directly measure, possibly as high as 240 km/s during the 18:39 UT window. With a termination shock needed to replicate the observed velocity gradients within the simulations, this provides strong empirical evidence for the existence of a termination shock. 

In addition to the Fe XXIV Doppler velocities observed by EIS, the non-thermal velocities also experience a sharp decline to a local minimum at the plasma-sheet base. 
Non-thermal velocities in this flare have been previously interpreted as a signal of macroscopic turbulence along the LOS, a key prediction of ongoing reconnection \citep{Warren2018,Longcope2018}. 
The simultaneous drop in the non-thermal velocities alongside the Doppler velocities provides further evidence of a termination shock front suppressing these flows. 
The non-thermal velocity variation along the current sheet has been reported from 3D flare models in the past \citep[e.g.,][]{Shen2023}. Models predict an increase in non-thermal velocity beneath the termination shock due to the enhanced plasma turbulence in the below-shock region, which is also consistent with observations (Figure \ref{fig:profile}, at heights below $\sim$1015 and $\sim$1025\arcsec\ at 16:44 and 18:39 UT respectively). However, the deduced non-thermal velocity from observed Fe XXIV might be affected by multiple-scale hot plasma flows along the LOS. Therefore, it is difficult to compare quantitatively in the current study. Further investigation of non-thermal velocities around the termination shock, and how the shock front affects turbulent flows, are left as an area of future study.

Finally, the spatial location of RHESSI 35-50 keV X-rays, at the location of minimum Doppler/non-thermal velocities within the flare cusp (high above the thermal 6--12 keV source located at the flare looptop), indicate trapped high-energy electrons above the flare looptop. This is another key prediction of a flare termination shock. Together, these observational constraints provide evidence for the presence of a termination shock within the September 10th 2017 solar flare.

Termination shocks can occur as long as super-magnetosonic reconnection outflows are present during the flare energy release processes. Given the magnitude and longevity of the September 10th 2017 flare, it is not necessarily surprising to see evidence of a termination shock this late into the flare's evolution. Previous works on this event have found other examples of high energy flare behavior several hours into the flare, including prolonged high plasma-sheet velocities/temperatures and growth of the hot loop-top cusp \citep{French2020}, prolonged plasma sheet downflows and microwave emission \citep{Yu2020}, and extended (12+ hours) gamma ray emission \citep{Omodei2018}. 

To our knowledge, the results in this paper present the first reported evidence of reconnection outflows driving a termination shock using direct Doppler shift measurements.

\section*{Data Availability}

Hinode/EIS, RHESSI and AIA data are publicly available to all. Data cubes of the two featured simulation frames will be provided via reasonable request to the corresponding author.

\section*{Acknowledgements}

Hinode is a Japanese mission developed and launched by ISAS/JAXA, with NAOJ as domestic partner and NASA and UKSA as international partners. It is operated by these agencies in co-operation with ESA and NSC (Norway). R.F. thanks support from the Brinson Prize Fellowship. S.J. and B.C. are supported by US National Science Foundation grant AGS-2108853 and NASA grant 80NSSC20K1318 to New Jersey Institute of Technology. C.S. is supported by NSF grant AST-2108438 and NASA grants 80NSSC21K2044, 80NSSC20K1318 to the Smithsonian Astrophysical Observatory. S.A.M. is supported by UKSA grant (ST/X002063/1) and STFC grant (ST/W001004/1).



\bibliographystyle{mnras}

\bsp	
\label{lastpage}
\end{document}